%% file: root.tex
\documentclass[letterpaper,10pt,conference]{ieeeconf}

\IEEEoverridecommandlockouts
\usepackage{graphicx}
\usepackage[cmex10]{amsmath}
\usepackage{subcaption}
\usepackage{booktabs}
\usepackage{siunitx}
\usepackage{epstopdf}
\usepackage{multirow}
\usepackage{amssymb}
\usepackage[table]{xcolor}
\usepackage{tikz}
\usepackage{myPlotStyle}
\usetikzlibrary{calc, positioning, decorations.pathreplacing, arrows.meta}
\usepackage{pgfplots}
\usepgfplotslibrary{fillbetween}
\usepackage{standalone}
\usepackage{dsfont}
\usepackage{url}

\pgfplotsset{
  split/.style={sBlue, mark=*, thick},
  crc/.style={sRed, mark=square*, thick},
  target/.style={sBlack, dashed},
  legend cell align=left,
}

\definecolor{darkred}{RGB}{180,0,0}

\newtheorem{theorem}{Theorem}
\newtheorem{remark}{Remark}
\newtheorem{proposition}{Proposition}

\definecolor{sGray}{RGB}{127,127,127}
\definecolor{sTeal}{RGB}{23,190,207}
\definecolor{sBlackColor}{RGB}{0,0,0}
\tikzset{sBlack/.style={color=sBlackColor}}
\newcommand{\bx}{\boldsymbol{x}}
\newcommand{\bu}{\boldsymbol{u}}
\newcommand{\by}{\boldsymbol{y}}
\newcommand{\bz}{\boldsymbol{z}}
\newcommand{\be}{\boldsymbol{e}}
\newcommand{\bff}{\boldsymbol{f}}
\newcommand{\bgg}{\boldsymbol{g}}
\newcommand{\bpsi}{\boldsymbol{\psi}}
\newcommand{\bzero}{\boldsymbol{0}}
\newcommand{\bchi}{\boldsymbol{\chi}}

\title{\LARGE \bf
Verification and Validation of Physics-Informed Surrogate Component Models for Dynamic Power-System Simulation
}

\author{
Petros Ellinas$^{1,2}$,
Indrajit Chaudhuri$^{2}$,
Johanna Vorwerk,
Spyros Chatzivasileiadis%
\thanks{$^{1}$Corresponding author: petrel@dtu.dk}%
\thanks{$^{2}$These authors contributed equally.}%
\thanks{All authors are with the Department of Wind and Energy Systems, Technical University of Denmark (DTU), Kgs.\ Lyngby, Denmark.}%
\thanks{This work was supported by the ERC Starting Grant VeriPhIED, funded by the European Research Council, Grant Agreement 949899.}
}

\begin{document}

\maketitle
\thispagestyle{empty}
\pagestyle{empty}

\begin{abstract}
Physics-informed machine learning surrogates are increasingly explored to accelerate dynamic simulation of generators, converters, and other power grid components. The key question, however, is not only whether a surrogate matches a stand-alone component model on average, but whether it remains accurate after insertion into a differential-algebraic simulator, where the surrogate outputs enter the algebraic equations coupling the component to the rest of the system. This paper formulates that in-simulator use as a verification and validation (V\&V) problem. A finite-horizon bound is derived that links allowable component-output error to algebraic-coupling sensitivity, dynamic error amplification, and the simulation horizon. Two complementary settings are then studied: model-based verification against a reference component solver, and data-based validation through conformal calibration of the component-output variables exchanged with the simulator. The framework is general, but the case study focuses on physics-informed neural-network surrogates of second-, fourth-, and sixth-order synchronous-machine models. Results show that good stand-alone surrogate accuracy does not by itself guarantee accurate in-simulator behavior, that the largest discrepancies concentrate in stressed operating regions, and that small equation residuals do not necessarily imply small state-trajectory errors.
\end{abstract}


\section{Introduction}

Dynamic simulation is a central tool for assessing disturbance propagation, controller performance, operational security, and real-time testing in modern power systems. The proliferation of converter-interfaced components such as batteries, solar PV, wind farms, and electric vehicles further increases model complexity and simulation cost \cite{kundur1994power,milano2020foundations,sun2023dynamic}.

A practical response is to replace selected dynamic \emph{component} models inside an existing simulator by surrogates. Here, we do not replace the full simulator; we retain the network equations, numerical workflow, and remaining component models, and only substitute selected, expensive to evaluate components. This preserves the trusted simulation structure while reducing the cost of repeatedly evaluating detailed component dynamics \cite{hamid2023deep,stiasny2023pinnsim}.

We use the term \emph{deployment} in this paper in that precise sense: a reference component model is replaced by a surrogate inside an existing dynamic simulator, while the rest of the simulator remains unchanged. The key question is then simple: \emph{when is a surrogate component model accurate enough for in-simulator use?} The answer cannot be based only on training loss or average simulation error over sampled trajectories. Once a surrogate is inserted into a simulator, its outputs are coupled to algebraic network equations and influence the trajectories of neighboring components. A locally small component error can therefore produce a non-negligible simulator-level deviation. The relevant notion is thus simulator-level accuracy, not only stand-alone component accuracy. The key quantities are the component outputs exchanged with the surrounding simulator through the algebraic equations, such as currents, voltages, or other coupling variables. In the remainder of the paper, we refer to these quantities as \emph{interface variables}.

Physics-informed machine learning (PIML) is an especially relevant instance of this broader surrogate-model class. By embedding governing differential equations in the training objective, PIML has emerged as a promising approach for approximating dynamic component behavior while preserving physical structure \cite{raissi2019physicsinformed,mao2020fpinns}. In power systems, recent work has shown that PIML surrogates can accelerate component simulation and can be integrated into bigger simulations \cite{stiasny2023pinnsim,ELLINAS2025101818}. In this paper, we use physics-informed neural-network surrogates of synchronous machines as the guiding example, but the framework applies more broadly to surrogate models of dynamic power-system components.

We consider two complementary settings. When a reference model or accepted high-fidelity solver is available, the problem is \emph{verification}: estimate the largest discrepancy over an admissible operating set and check whether it satisfies an in-simulator accuracy requirement. When only playback data or measurements are available, the problem is \emph{validation}: quantify uncertainty on the component-output variables that the simulator actually uses. The tools used in the two settings are established: sensitivity-based methods from power-system analysis can be used when a reference model is available, while uncertainty calibration and black-box search can be used when only playback data or expensive evaluations are available \cite{HiskensPai2000,fourati25ecp,linderman_conformal}. The contribution of this paper is to cast surrogate deployment as a simulator-level V\&V problem for differential-algebraic simulation (DAE), and to organize these tools around a finite-horizon acceptance criterion for in-simulator use.

The first technical step is to connect component-level error to simulator-level deviation. We derive a finite-horizon bound that links allowable component-output error to the sensitivity of the algebraic coupling, the amplification of errors by the differential dynamics, and the simulation horizon. This gives a concrete acceptance threshold for in-simulator use. On top of that formulation, differentiable worst-case search becomes a natural tool for verification when the reference solver is differentiable, and conformal calibration becomes a natural tool for validation when only playback data are available.

The main contributions are:
\begin{enumerate}
    \item \textbf{Simulator-level acceptance condition:} we formulate surrogate deployment as an in-simulator specification problem and derive a finite-horizon bound that links allowable component-output error to algebraic-coupling sensitivity, dynamic error amplification, and the simulation horizon.
    \item \textbf{V\&V methodology:} we combine differentiable worst-case search of surrogate discrepancy with conformal calibration of interface variables, covering both model-based verification and data-based validation.
    \item \textbf{Physics-informed case study:} we demonstrate the framework on second-, fourth-, and sixth-order synchronous-machine surrogates and identify failure modes that are not visible from average-case metrics alone. Code available at: \url{https://github.com/elpetros99/PINNProof}.
\end{enumerate}
The remainder of the paper is organized as follows. Section~\ref{sec:operator}
introduces surrogate component models and distinguishes the error notions relevant
for deployment. Section~\ref{sec:closedloop} formulates the in-simulator certification
problem and derives the finite-horizon acceptance bound. Section~\ref{sec:methods}
presents the proposed verification and validation methods. Section~\ref{sec:results}
reports the case-study results for synchronous-machine surrogates. Finally,
Section~\ref{sec:conclusion} concludes the paper and discusses future work.

\section{Surrogate Component Models and Error Notions}
\label{sec:operator}

Consider a dynamic component model
\begin{equation}
\label{eq:ode_system}
\frac{d\bx(t)}{dt}
=
\bff\bigl(\bx(t),\bu(t),t\bigr),
\qquad
\bx(t_0)=\bx_0,
\end{equation}
where $\bx(t)\in\mathbb{R}^n$ denotes the differential state and $\bu(t)\in\mathbb{R}^r$ denotes exogenous inputs or control variables. Let
$G(\bx_0,\bu(\cdot),t_0,\tau)
=
\bx(t_0+\tau)$
denote the reference component solution map. A surrogate component model is any map
$U(\bx_0,\bu(\cdot),t_0,\tau)
\approx
G(\bx_0,\bu(\cdot),t_0,\tau),$
designed to approximate the same input-output relation more efficiently.

In this paper, the main instantiation is a PIML surrogate $U_\theta$, parametrized by $\theta$, for which
$U_\theta(\bx_0,\bu(\cdot),\tau)\approx \bx(\tau)$,
and the training loss combines residual consistency, data fit, and initial-condition consistency:
\begin{equation}
\label{eq:piml_loss}
\begin{aligned}
\mathcal{L}(\theta)
&=
\frac{1}{N_r}\sum_{i=1}^{N_r}
\left\|
\frac{\partial U_\theta}{\partial \tau}
-
\bff\!\left(U_\theta,\bu_i(\tau_i),\tau_i\right)
\right\|_2^2\\
&\quad+
\frac{1}{N_d}\sum_{i=1}^{N_d}
\left\|
U_\theta(\bx_{0,i},\bu_i,\tau_i)-\bx_i
\right\|_2^2\\
&\quad+
\frac{1}{N_0}\sum_{i=1}^{N_0}
\left\|
U_\theta(\bx_{0,i},\bu_i,0)-\bx_{0,i}
\right\|_2^2 .
\end{aligned}
\end{equation}

Two error notions are central. The \emph{functional} (training) error is the ODE residual along the surrogate trajectory.
\begin{equation}
\label{eq:functional_error}
e_{\mathrm{fun}}(\bx_0,\bu,\tau)
=
\left\|
\frac{\partial U_\theta}{\partial \tau}
-
\bff\!\left(U_\theta,\bu(\tau),\tau\right)
\right\|_2,
\end{equation}
whereas the \emph{solution error} is the discrepancy between the surrogate state and the reference trajectory,
\begin{equation}
\label{eq:solution_error}
e_{\mathrm{sol}}(\bx_0,\bu,\tau)
=
\left\|
U_\theta(\bx_0,\bu,\tau)
-
G(\bx_0,\bu,\tau)
\right\|_2.
\end{equation}
The distinction matters because residual consistency is useful for training, but in-simulator acceptability is governed by the solution error and the error in the component outputs exchanged with the simulator.

\section{In-Simulator Certification Problem}
\label{sec:closedloop}

Let $G$ denote a reference component map and let $U$ denote a surrogate component model. Over an admissible set $\mathcal{U}$ of initial conditions, inputs, and horizons, we define the worst-case \emph{state} discrepancy
\begin{equation}
\label{eq:ver_opt}
E_x(\mathcal{U})
=
\sup_{(\bx_0,\bu(\cdot),\tau)\in\mathcal{U}}
\left\|
G(\bx_0,\bu(\cdot),\tau)
-
U(\bx_0,\bu(\cdot),\tau)
\right\|,
\end{equation}
and the induced worst-case discrepancy in the component outputs exchanged with the simulator, for a chosen map $\bpsi$,
\begin{equation}
\label{eq:ver_opt_z}
E_z(\mathcal{U})
=
\sup_{(\bx_0,\bu(\cdot),t)\in\mathcal{U}}
\left\|
\bpsi(\bx(t),\by(t),\bu(t))
-
\bpsi(\hat\bx(t),\hat\by(t),\bu(t))
\right\|.
\end{equation}
Verification and validation are then posed as a specification problem:
\begin{equation}
\label{eq:spec_problem}
R\!\left(E_z(\mathcal{U})\right)\le \varepsilon,
\end{equation}
where $R(\cdot)$ encodes the downstream simulator requirement. The next subsection gives an explicit and sufficient choice of $R$.

\subsection{Embedding a component surrogate in a differential-algebraic simulator}

Consider a power-system DAE
\begin{align}
\dot{\bx} &= \bff(\bx,\by,\bu), \label{eq:cl_dae_x}\\
\bzero &= \bgg(\bx,\by,\bz,\bu), \label{eq:cl_dae_alg}\\
\bz &= \bpsi(\bx,\by,\bu), \label{eq:cl_true_interface}
\end{align}
where $\bx\in\mathbb{R}^n$ are differential states, $\by\in\mathbb{R}^m$ are algebraic variables,
$\bu\in\mathbb{R}^r$ are exogenous inputs or controls, and $\bz\in\mathbb{R}^p$ collects the component outputs exchanged with the rest of the simulator through the algebraic equations. We refer to $\bz$ as the interface signal.

In the surrogate setting, the reference component trajectory $(\bx(t),\by(t))$ is replaced by an approximation $(\hat\bx(t),\hat\by(t))$, while the simulator computes the interface through the same map:
\begin{equation}
\label{eq:cl_surrogate_interface}
\hat{\bz}=\bpsi(\hat{\bx},\hat{\by},\bu).
\end{equation}
Define the embedded trajectory errors
\begin{equation}
\label{eq:state_errors}
\be_x=\hat{\bx}-\bx,\qquad
\be_y=\hat{\by}-\by,\qquad
\be_z=\hat{\bz}-\bz.
\end{equation}
The acceptance question is whether a uniform interface bound
\begin{equation}
\label{eq:interface_bound}
\|\be_z(t)\|\le \varepsilon,\qquad \forall t\in[0,T],
\end{equation}
is sufficient to guarantee an acceptable simulator-level deviation.

\subsection{Finite-horizon in-simulator bound}
Assume that the algebraic map induced by \eqref{eq:cl_dae_alg} is globally Lipschitz in $\bx$ and $\bz$ over the admissible operating region. Then
\begin{equation}
\label{eq:ey_bound}
\|\be_y(t)\|
\le
K_{yz}\|\be_z(t)\|
+
K_{yx}\|\be_x(t)\|,
\end{equation}
where $K_{yz}$ and $K_{yx}$ are the corresponding Lipschitz constants. These constants quantify how strongly the algebraic equations react to perturbations in the component outputs and states over the chosen operating region. Practically, they can be estimated from local algebraic Jacobians or conservatively from numerical perturbation studies over the operating region. Assume further that $\bff$ is Lipschitz in $\by$ with constant $L_y$ and one-sided Lipschitz in $\bx$ with constant $\mu_{cl}$ over the same region. Together, these quantities summarize how errors introduced by the surrogate are transmitted through the algebraic coupling and amplified by the differential dynamics. Then
\begin{equation}
\label{eq:ex_bound_pre}
\frac{d}{dt}\|\be_x\|
\le
\mu_{cl}\|\be_x\|+L_y\|\be_y\|.
\end{equation}
Combining \eqref{eq:ey_bound} with \eqref{eq:interface_bound} gives
\begin{equation}
\label{eq:ex_bound_mid}
\frac{d}{dt}\|\be_x\|
\le
(\mu_{cl}+L_yK_{yx})\|\be_x\|
+
L_yK_{yz}\varepsilon.
\end{equation}
Let
\begin{equation}
\alpha:=\mu_{cl}+L_yK_{yx},
\qquad
\Phi(T,\alpha)=
\begin{cases}
\dfrac{e^{\alpha T}-1}{\alpha}, & \alpha\neq 0,\\[4pt]
T, & \alpha=0.
\end{cases}
\end{equation}

\begin{theorem}[Finite-horizon in-simulator acceptance bound]
\label{thm:closedloop}
Assume the Lipschitz and one-sided Lipschitz conditions above, and assume that the reference and surrogate-embedded trajectories share the same initial condition, so that $\be_x(0)=\be_y(0)=\bzero$. If
\begin{equation}
\|\be_z(t)\|\le \varepsilon,\qquad \forall t\in[0,T],
\end{equation}
then
\begin{equation}
\label{eq:thm_ex}
\|\be_x(T)\|
\le
L_yK_{yz}\Phi(T,\alpha)\,\varepsilon,
\end{equation}
\begin{equation}
\label{eq:thm_ey}
\|\be_y(T)\|
\le
K_{yx}L_yK_{yz}\Phi(T,\alpha)\,\varepsilon
+
K_{yz}\varepsilon,
\end{equation}
and therefore
\begin{equation}
\label{eq:thm_total}
\|\be_x(T)\|+\|\be_y(T)\|
\le
\Bigl[(1+K_{yx})L_yK_{yz}\Phi(T,\alpha)+K_{yz}\Bigr]\varepsilon.
\end{equation}
Hence, if $\Delta>0$ is the maximum admissible simulator-level deviation, then a sufficient interface requirement is
\begin{equation}
\label{eq:eps_max}
\varepsilon\le
\varepsilon_{\max}
=
\frac{\Delta}
{(1+K_{yx})L_yK_{yz}\Phi(T,\alpha)+K_{yz}}.
\end{equation}
\end{theorem}

\begin{remark}
Theorem~\ref{thm:closedloop} can be used as a practical acceptance template. For a chosen operating region and simulation horizon, the quantities $K_{yz}$, $K_{yx}$, $L_y$, and $\mu_{cl}$ can be estimated from the reference model using local Jacobians, sensitivity calculations, or conservative numerical perturbation studies. These estimates then yield an admissible output-error threshold $\varepsilon_{\max}$ for deciding whether a surrogate is suitable for in-simulator use over that region and horizon.
\end{remark}

\section{Verification and Validation Methods}
\label{sec:methods}

The previous section identifies what must be controlled for in-simulator use: the interface error. In practice, the bound is used in two steps: first estimate or conservatively upper-bound the quantities appearing in Theorem~\ref{thm:closedloop} over the operating region of interest, and then check whether the surrogate discrepancy found by verification or validation remains below the resulting threshold $\varepsilon_{\max}$. The remaining question is computational: how do we estimate whether $E_z(\mathcal{U})\le \varepsilon_{\max}$?

\subsection{Differentiable worst-case verification}

When a reference model or accepted solver is available, we seek the largest discrepancy over the admissible operating set:
\begin{equation}
\label{eq:verification_objective}
\max_{\eta\in\Omega}
\mathcal{J}(\eta)
:=
\max_{t\in[0,T]}
\left\|
U(\eta,t)-G(\eta,t)
\right\|_2,
\end{equation}
where $\eta$ collects optimized variables such as initial conditions and disturbance parameters. In the differentiable setting, the objective is maximized using gradients obtained through the reference solver or through adjoint/trajectory-sensitivity machinery \cite{HiskensPai2000}. We refer to this as \emph{differentiable verification}. Depending on the experiment, $\mathcal{J}$ can be defined on a state discrepancy or on an interface discrepancy; in the results below, Fig.~\ref{fig:methods_comp} reports state-discrepancy maximization, while the SMIB test separately reports interface and simulator-level deviations.

To complement worst-case search, we use novelty-based trajectory generation as a coverage-oriented evaluation tool. Let $\tau(\eta)$ denote the trajectory associated with $\eta$. Given previously sampled trajectories $\{\tau_1,\dots,\tau_k\}$, define
\begin{equation}
\label{eq:novelty_score}
\mathcal{N}(\eta)
=
-\frac{1}{\beta}
\log
\sum_{i=1}^{k}
\exp\!\left(-\beta\,d\bigl(\tau(\eta),\tau_i\bigr)^2\right),
\end{equation}
where $d$ is a trajectory metric and $\beta>0$ is a sharpness parameter. Maximizing \eqref{eq:novelty_score} promotes coverage of distinct dynamical behaviors rather than direct maximization of surrogate error.

For non-differentiable settings, including hardware-in-the-loop or black-box reference models, we use ECP \cite{fourati25ecp} to concentrate expensive evaluations on regions that may still contain larger discrepancies than the current best.

\subsection{Statistical validation of component-output variables}

When only playback data or measurements are available, deterministic worst-case optimization is generally unavailable. We then validate the surrogate on a component-output variable $r$ used by the simulator (for example, a current channel) using features $\bchi$ built from available simulator signals. Let $\{(\bchi_i,r_i)\}_{i=1}^{n}$ be a calibration set and $\hat r_i=h_\theta(\bchi_i)$ the surrogate prediction. Define the nonconformity score
\begin{equation}
\label{eq:conformal_score}
v_i=\frac{|r_i-\hat r_i|}{\sigma(\bchi_i)},
\end{equation}
where $\sigma(\bchi)>0$ is a scale function. Split conformal then yields
\begin{equation}
\label{eq:split_cp}
C_\alpha(\bchi)
=
\left[
\hat r(\bchi)-q_\alpha \sigma(\bchi),\,
\hat r(\bchi)+q_\alpha \sigma(\bchi)
\right],
\end{equation}
with marginal guarantee $\mathbb{P}\{r\in C_\alpha(\bchi)\}\ge 1-\alpha$ under exchangeability.

These are precisely the variables that enter the algebraic coupling; following the notation above, we refer to them as interface variables. We also use a high-confidence UCB variant and connect any resulting interface bound to Theorem~\ref{thm:closedloop}:

\begin{proposition}
\label{prop:cp_to_closedloop}
Suppose calibration yields a uniform interface bound $\|\be_z(t)\|\le \bar\varepsilon$ for all $t\in[0,T]$. Under the assumptions of Theorem~\ref{thm:closedloop},
\begin{equation}
\|\be_x(T)\|+\|\be_y(T)\|
\le
\Bigl[(1+K_{yx})L_yK_{yz}\Phi(T,\alpha)+K_{yz}\Bigr]\bar\varepsilon.
\end{equation}
\end{proposition}

\section{Results}
\label{sec:results}

The results are organized around the paper's central question: \emph{when is a surrogate component model accurate enough for in-simulator use?} We first verify the trend predicted by Theorem~\ref{thm:closedloop} on a simple single-machine infinite-bus benchmark, and then study empirical worst-case errors, residual-versus-solution error, and interface-level validation for PIML synchronous-machine surrogates.

\subsection{Benchmark models and setup}

We study PIML surrogates of three synchronous-machine component models of increasing complexity: the two-state swing model (SM2), the fourth-order model (SM4), and the sixth-order model (SM6). Parameters and training setup follow \cite{ELLINAS2025101818}. All surrogates use three hidden layers with 64 neurons per layer and are trained with L-BFGS. Unless stated otherwise, the training horizon is $t_{\max}=0.2\,\mathrm{s}$.

The validation methodology is intended for playback, measurement, or experimental settings \cite{Lu2014,IEEE1646-2004}. In the present paper, the reported validation experiments use high-fidelity simulated playback trajectories.

All worst-case search methods use the same box constraints on the optimized variables and matched evaluation budgets: $10$ random restarts and $100$ total trajectory evaluations per model. Thus, all reported worst-case values are \emph{empirical maxima found under a fixed budget}, not formal global certificates. Novelty-based sampling uses the same total budget as naive sampling and the metric
\begin{equation}
d(\tau_a,\tau_b)^2=\frac{1}{T}\int_0^T \|\bx_a(t)-\bx_b(t)\|_2^2\,dt,
\end{equation}
computed on the same time grid as the simulator outputs.

\begin{figure}[t]
    \centering

    \begin{subfigure}[t]{0.48\columnwidth}
        \centering
        \includegraphics[width=\linewidth]{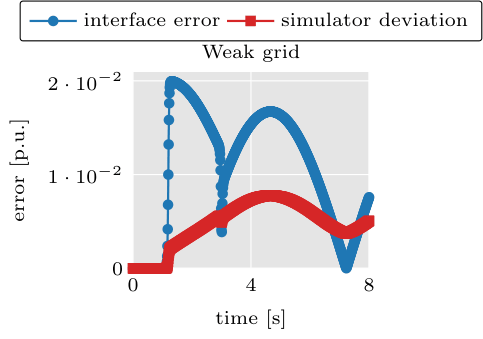}
    \end{subfigure}
    \hfill
    \begin{subfigure}[t]{0.48\columnwidth}
        \centering
        \includegraphics[width=\linewidth]{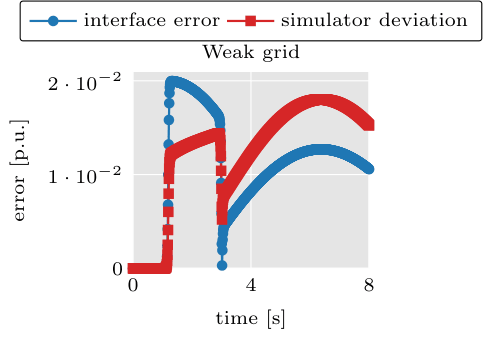}
    \end{subfigure}

    \vspace{0.4em}

    \makebox[\columnwidth]{%
        \begin{subfigure}[t]{0.48\columnwidth}
            \centering
            \includegraphics[width=\linewidth]{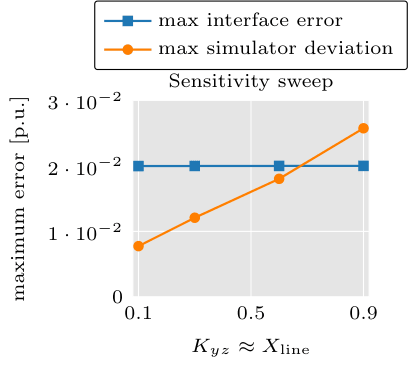}
        \end{subfigure}
    }

    \caption{Illustration of Theorem~\ref{thm:closedloop} on a single-machine infinite-bus (SMIB) system. Top: for nearly identical maximum interface errors $\|\be_z(t)\|$, the simulator deviation $\|\be_x(t)\|+\|\be_y(t)\|$ is larger in the weak-grid case than in the strong-grid case. Bottom: at fixed target $\varepsilon$, the maximum simulator deviation increases with grid sensitivity ($K_{yz}\approx X_{\mathrm{line}}$).}
    \label{fig:smib_theorem_validation}
\end{figure}





We first isolate the theorem's main mechanism on a standard single-machine infinite-bus (SMIB) system, consisting of a synchronous machine connected to an infinite bus through a transmission line. In the notation of Theorem~\ref{thm:closedloop}, $\bx=[\delta,\omega]^\top$, $\by=V\in\mathbb{C}$ is the terminal-voltage phasor, and $\bz=I\in\mathbb{C}$ is the terminal injected-current phasor. Varying the line reactance $X_{\mathrm{line}}$ changes the sensitivity of the algebraic coupling while keeping the component dynamics fixed.

To avoid conflating learning error with network amplification, this experiment does not use a learned surrogate. Instead, we perturb the current interface directly and study how the same interface discrepancy propagates through grids of different strength. Let $\bar E(\delta)=E'e^{j\delta}$ and $X_{\mathrm{eq}}=X_d'+X_{\mathrm{line}}$. The network equations are
\begin{equation}
I=\frac{\bar E(\delta)-V_\infty}{jX_{\mathrm{eq}}},\qquad
V=V_\infty + jX_{\mathrm{line}} I,
\end{equation}
with $P_e=\Re(V\,\overline{I})$ and swing dynamics $\dot\delta=\omega$, $2H\dot\omega=P_m(t)-P_e-D\omega$ under a step $P_m(t)=P_{m0}+dP_m\,\mathbf{1}\{t\ge t_{\mathrm{step}}\}$. We use $H{=}3.5$, $D{=}0$, $E'{=}1.1$, $X_d'{=}0.3$, $P_{m0}{=}0.7$, $dP_m{=}0.08$, $t_{\mathrm{step}}{=}1.0$\,s, $T{=}8$\,s, and RK4 with $\Delta t=0.01$\,s.

The perturbed interface is
\begin{equation}
\hat I = I(\hat\delta)+d(t),\qquad d(t)=a\,s(t)\,e^{j\phi},
\end{equation}
where $s(t)$ is the windowed $\tanh$ function used in the script and $(t_{\mathrm{on}},t_{\mathrm{off}},w,\phi)=(1.2,3.0,0.03,0.7)$. The amplitude is chosen from a target interface budget $\varepsilon$ via $a=\varepsilon\frac{X_d'+X_{\mathrm{line}}}{X_d'}$, so that $\max_t\|\be_z(t)\|=\max_t|\hat I(t)-I(t)|\approx\varepsilon$ across different $X_{\mathrm{line}}$ values. We then evaluate the simulator-deviation metric
\begin{equation}
e_{\mathrm{sim}}(t):=\|\be_x(t)\|_2+\|\be_y(t)\|,\qquad
\be_y(t)=\hat V(t)-V(t).
\end{equation}

The result is the central intuition behind the paper. With $\varepsilon=0.02$, the maximum interface discrepancy is essentially the same in the strong-grid and weak-grid cases ($0.01995$ versus $0.01997$), yet the maximum simulator deviation increases from $0.00775$ to $0.01801$, i.e., by approximately $2.3\times$. Thus, the same component-level interface error can be acceptable in one network and problematic in another. The lower panel of Fig.~\ref{fig:smib_theorem_validation} shows the same monotone trend over a sweep of $X_{\mathrm{line}}$, consistent with Theorem~\ref{thm:closedloop}: simulator-level suitability depends not only on surrogate accuracy, but also on the sensitivity of the surrounding network.

\subsection{Empirical worst-case verification of PIML surrogates}

\begin{figure}[t]
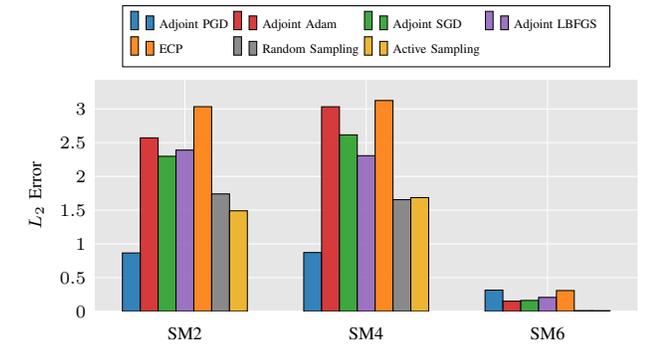
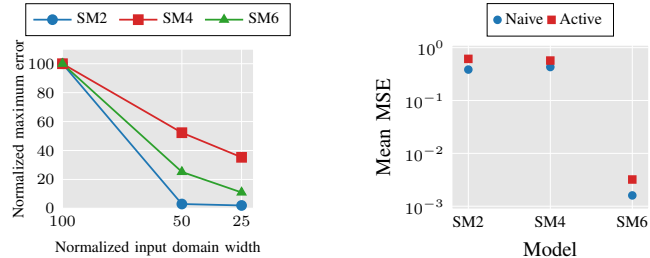

  \centering
  \begin{subfigure}{\columnwidth}
    \centering
    \includestandalone[width=0.95\columnwidth]{figures/Methods_comparison}
    \caption{Comparison of differentiable search (Adjoint PGD/Adam/SGD/LBFGS), ECP, and sampling-based baselines.}
    \label{fig:methods_comp}
  \end{subfigure}

  \vspace{0.5em}

  \begin{subfigure}[t]{0.44\columnwidth}
    \centering
    \includestandalone[width=\columnwidth]{figures/methods_reducing_bounds}
    \caption{Normalized maximum error versus normalized search-box width.}
    \label{fig:methods_bounds}
  \end{subfigure}
  \hfill
  \begin{subfigure}[t]{0.44\columnwidth}
    \centering
    \includestandalone[width=\columnwidth]{figures/sampling_methods_mean_mse}
    \caption{Mean MSE for naive and novelty-based sampling.}
    \label{fig:sampling_mse}
  \end{subfigure}

  \caption{Worst-case verification and coverage-oriented sampling under matched budgets.}
  \label{fig:all_methods}
\end{figure}

We next study the PIML surrogates directly. The goal here is to identify where the largest discrepancies occur over the admissible operating set and to assess surrogate behavior over a dynamically diverse set of trajectories. Figure~\ref{fig:methods_comp} compares differentiable worst-case search, ECP, and sampling-based baselines under matched budgets. The bars labeled Adjoint PGD, Adjoint Adam, Adjoint SGD, and Adjoint LBFGS correspond to \emph{differentiable verification}: the discrepancy objective is maximized using gradients through the reference solver or its sensitivity machinery.

The main result is consistent across the tested models: verification-oriented search identifies larger discrepancies than average sampling. This indicates that the worst surrogate errors do not occur in the bulk of randomly sampled operating points, but in a smaller set of more adverse scenarios that average-case evaluation can miss. A surrogate that appears accurate during random evaluations can still exhibit significantly larger errors in specific operating regions. The fact that both ECP and adjoint-based searches outperform the sampling baselines also suggests a nonconvex error landscape in which global black-box search can complement gradient-based optimization.

Figure~\ref{fig:methods_bounds} provides further intuition about where these errors occur. As the admissible search box is reduced, the empirical worst-case discrepancy decreases substantially. This suggests that the largest errors arise near the boundaries of the operating region rather than near nominal conditions. A plausible explanation is that these edge cases contain less uniform dynamics, with some variables evolving faster than others, which makes them harder to learn accurately with gradient-based training. Thus, surrogate accuracy varies across the operating region, with smaller discrepancies near nominal conditions and larger ones toward the boundaries.

Novelty-based trajectory generation is used to construct dynamically diverse test sets, and the mean-error comparison is reported in Fig.~\ref{fig:sampling_mse}. Compared with naive sampling, novelty sampling evaluates the surrogate over a broader and more dynamically diverse set of trajectories within the admissible operating region. This makes it useful for assessing surrogate behavior across the range for which it is intended, rather than for finding the largest discrepancy directly. As expected, the largest discrepancies are still found by explicit worst-case search, since novelty sampling targets coverage rather than direct maximization of error.

\subsection{Residual error, solution error, and interface-level validation}

\begin{figure}[t]
    \centering
    \begin{subfigure}{\linewidth}
        \centering
        \includegraphics[width=\linewidth]{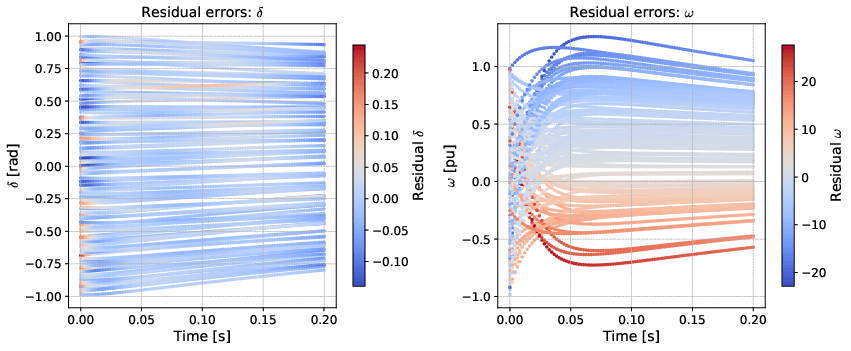}
        \caption{Functional (residual) error for $\delta$ and $\omega$.}
        \label{fig:residuals_delta_omega}
    \end{subfigure}

    \vspace{0.10cm}

    \begin{subfigure}{\linewidth}
        \centering
        \includegraphics[width=0.95\linewidth]{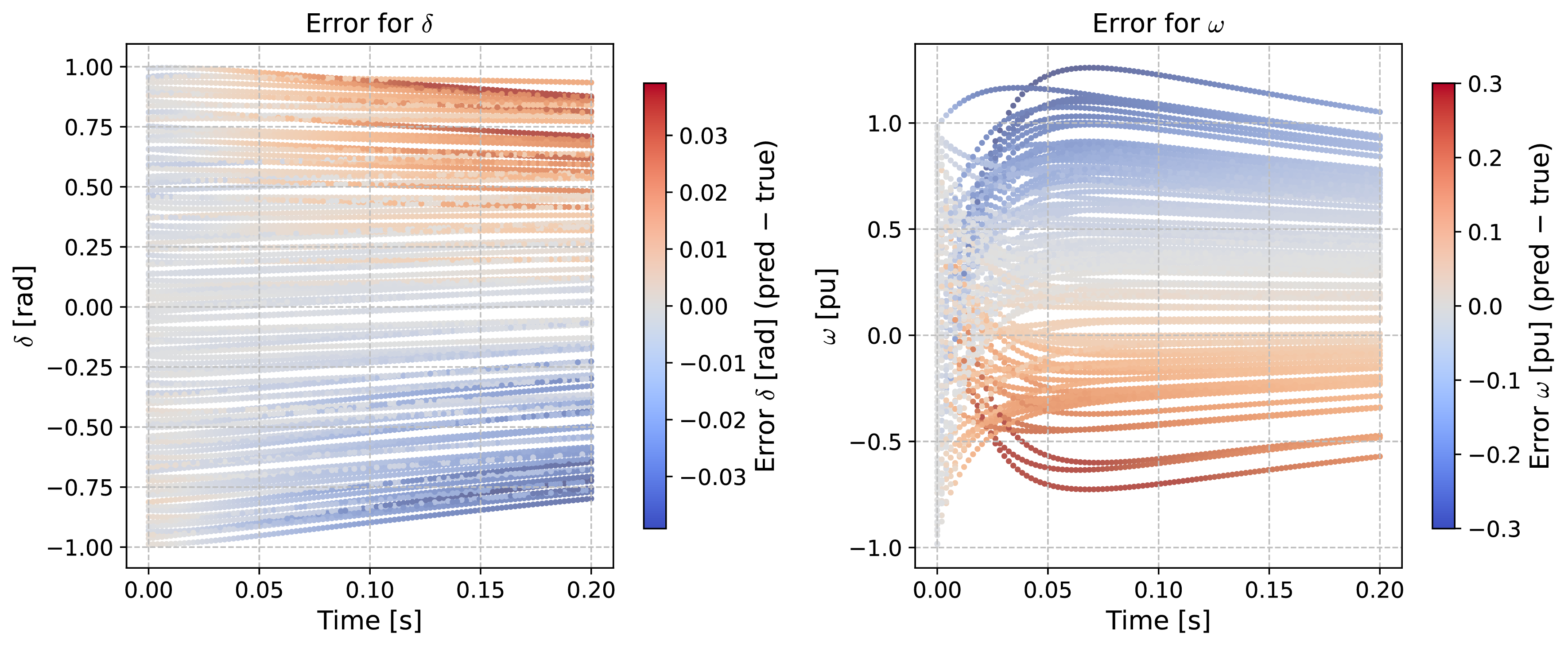}
        \caption{Solution error for $\delta$ and $\omega$.}
        \label{fig:solution_error_delta_omega}
    \end{subfigure}

    \caption{Functional and solution errors are related but not equivalent.}
    \label{fig:delta_omega_errors}
\end{figure}

\begin{figure}[t]
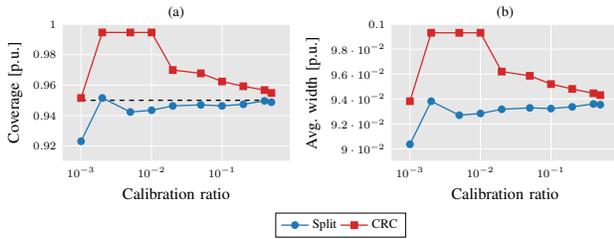

  \centering
  \includestandalone[width=0.95\linewidth]{figures/conformal_results}
  \caption{Split conformal versus high-confidence UCB-based calibration across calibration ratios.}
  \label{fig:calibration_results_main}
\end{figure}

Figure~\ref{fig:delta_omega_errors} compares residual-based functional error with solution error for SM6. The two are clearly related, but they are not the same object and they evolve differently in time. The functional error appears in localized bursts, especially around rapid transients, whereas the solution error accumulates over time and can produce visible drift even when the residual is not large at every instant.

This distinction matters for simulation use. Residual minimization is useful during training, but the surrounding simulator does not directly ``see'' the residual; it sees the state and interface variables produced by the surrogate. Thus, a small residual does not by itself guarantee a small simulator-level deviation. This is especially relevant in boundary regions of the operating set, where the dynamics are less uniform and some variables evolve faster than others. Such mixed fast--slow behavior is harder to learn uniformly with gradient-based training and can therefore lead to accumulated mismatch in the interface variables even when the instantaneous residual remains moderate.

The interface-level analysis also indicates where these effects matter most. In the tested trajectories, the largest mismatch appears in $I_q$, the quadrature-axis current obtained after the rotational Park/$dq$ transformation. This is relevant because rotation-based interface mappings can amplify modest state errors: a small error in an angle-like variable can rotate the projected quantities and lead to a larger discrepancy in the exchanged interface channels. The key deployment question is therefore not whether every internal state is accurate in isolation, but whether the interface quantities seen by the surrounding simulator remain within an acceptable tolerance.

We therefore validate at the interface-variable level. Figure~\ref{fig:calibration_results_main} reports split conformal and a more conservative UCB-based calibration for $I_q$ using calibration ratio $\rho$ and nominal coverage $1-\alpha=0.95$. Standard split conformal tends to under-cover when the calibration set is small, whereas the UCB-based rule is wider but more reliable in the low-data regime. This is the relevant practical trade-off if the calibrated interval is to be interpreted as an operational bound on interface error.

\section{Conclusion}
\label{sec:conclusion}

This paper presented a V\&V framework for surrogate component models embedded in dynamic power-system simulation. The main result is a finite-horizon bound linking allowable component-output error to algebraic-coupling sensitivity, error growth, and the simulation horizon; the SMIB example empirically validates the trend predicted by this bound. In addition, differentiable worst-case search, novelty-based generation of dynamically diverse test trajectories, and conformal prediction for data-based validation of surrogate interface variables provide practical tools for assessing whether a surrogate is accurate enough for in-simulator use.

Applied to PIML synchronous-machine surrogates, the results show that good stand-alone accuracy does not by itself guarantee good simulator-level behavior, that the largest discrepancies concentrate near the boundaries of the operating region, and that small equation mismatches do not always imply small simulation errors. The experiments also highlight the value of evaluating surrogates on dynamically diverse trajectories rather than only nominal or randomly sampled cases. Overall, the present evidence supports these models as promising candidates for restricted operating regions, but not yet as unrestricted drop-in replacements. Future work will study tighter estimation of the acceptance-bound quantities and their use in practical certification workflows.



\smallskip
\noindent\textit{During the preparation of this work, the author, Petros Ellinas, used ChatGPT (OpenAI) to improve readability and check grammar. After using this tool, the author reviewed and edited the content as needed and takes full responsibility for the content of the published article.}

\bibliographystyle{IEEEtran}
\bibliography{IEEEabrv,ref}

\end{document}

%% file: figures/Methods_comparison.tex
\begin{tikzpicture}
\small
\begin{axis}[
    ybar = 0pt,
    width=7.5cm, height=3.2cm, 
    bar width=7pt,
    ymin=0,
    ylabel={$L_2$ Error},
    symbolic x coords={SM2,SM4,SM6},
    xtick=data,
    ymajorgrids,
    label style={font=\scriptsize},
    tick label style={font=\scriptsize},
    legend style={
        font=\tiny, 
        at={(0.5,1.05)}, 
        anchor=south, 
        legend columns=4,
        cells={anchor=west}
    },
    enlarge x limits=0.25,
    xlabel={},
    every axis plot/.append style={line width=0.4pt, fill opacity=0.9},
    every node near coord/.style={
        /pgf/number format/1000 sep=,
        font=\tiny, anchor=south, yshift=2pt
    },
]

\addplot+[black, fill=sBlue]   coordinates {(SM2,0.8652) (SM4,0.8716) (SM6,0.3145)};
\addlegendentry{Adjoint PGD}

\addplot+[black, fill=sRed]    coordinates {(SM2,2.5709) (SM4,3.0316) (SM6,0.1509)};
\addlegendentry{Adjoint Adam}

\addplot+[black, fill=sGreen]  coordinates {(SM2,2.2979) (SM4,2.6154) (SM6,0.1616)};
\addlegendentry{Adjoint SGD}

\addplot+[black, fill=sPurple] coordinates {(SM2,2.3897) (SM4,2.3065) (SM6,0.2068)};
\addlegendentry{Adjoint LBFGS}

\addplot+[black, fill=sOrange] coordinates {(SM2,3.0331) (SM4,3.1246) (SM6,0.3081)};
\addlegendentry{ECP}

\addplot+[black, fill=sGray]   coordinates {(SM2,1.74) (SM4,1.655) (SM6,0.0107)};
\addlegendentry{Random Sampling}

\addplot+[black, fill=sYellow] coordinates {(SM2,1.49) (SM4,1.6856) (SM6,0.0101)};
\addlegendentry{Active Sampling}

\end{axis}
\end{tikzpicture}

%% file: figures/methods_reducing_bounds.tex
\begin{tikzpicture}
\begin{axis}[
  width=3cm, height=2.5cm,
  xlabel={Normalized input domain width},
  ylabel={Normalized maximum error},
  xmin=20, xmax=100,
  ymin=0, ymax=110,
  x dir=reverse, 
  xtick={100,50,25},
  ytick={0,20,40,60,80,100},
  grid=both,
  legend style={
  at={(0.5,1.08)},
  anchor=south,
  legend columns=-1,
  /tikz/every even column/.append style={column sep=0.1cm}
},
   label style={font=\scriptsize},     
  tick label style={font=\scriptsize}, 
  legend style={font=\scriptsize}, 
  ylabel style={yshift=-0.5em}
]

\addplot+[mark=*, thick, sBlue] coordinates {(100,100) (50,3) (25,2)};
\addlegendentry{SM2}

\addplot+[mark=square*, thick, sRed] coordinates {(100,100) (50,52.3) (25,35.3)};
\addlegendentry{SM4}

\addplot+[mark=triangle*, thick,sGreen] coordinates {(100,100) (50,25.2) (25,11)};
\addlegendentry{SM6}

\end{axis}
\end{tikzpicture}

%% file: figures/sampling_methods_mean_mse.tex
\begin{tikzpicture}
\small
\begin{axis}[
  width=3cm, height=2.5cm,
  ymode=log,
  log basis y={10},
  ylabel={Mean MSE},
  xlabel={Model},
  symbolic x coords={SM2,SM4,SM6},
  xtick=data,
  ymajorgrids,
  xticklabel style={font=\scriptsize},
  legend style={font=\scriptsize, at={(0.5,1.05)}, anchor=south, legend columns=2},
  every axis plot/.append style={mark size=1.5pt, line width=0.7pt},
  scatter/classes={
    Naive={mark=*, sBlue},
    Active={mark=square*, sRed}
  },
]

\addplot[scatter,only marks,scatter src=explicit symbolic]
coordinates {(SM2,0.385)[Naive] (SM4,0.43)[Naive] (SM6,0.0016)[Naive]};
\addlegendentry{Naive}

\addplot[scatter,only marks,scatter src=explicit symbolic]
coordinates {(SM2,0.6133)[Active] (SM4,0.57)[Active] (SM6,0.0032)[Active]};
\addlegendentry{Active}

\end{axis}
\end{tikzpicture}

%% file: figures/conformal_results.tex
\begin{tikzpicture}
\pgfplotsset{
  split/.style={sBlue, mark=*, thick},
  crc/.style={sRed, mark=square*, thick},
  target/.style={sBlack, dashed, thick},
  legend cell align=left,
}
\begin{groupplot}[
  group style={group size=2 by 1, horizontal sep=2.2cm},
  width=5cm, height=3cm,
  xmode=log, log basis x=10,
  legend to name=grouplegend,
  legend columns=3
]

\nextgroupplot[
  xlabel={Calibration ratio},
  ylabel={Coverage [p.u.]},
  ymin=0.91, ymax=1.0,
  title={(a)}
]
\addplot[target, domain=0.001:0.5, samples=50] {0.95};
\addlegendentry{Target $1-\alpha=0.95$}


\addplot+[split] coordinates {
  (0.001,0.923051) (0.002,0.951546) (0.005,0.942316) (0.01,0.943437)
  (0.02,0.946325) (0.05,0.946996) (0.1,0.946326) (0.2,0.947357)
  (0.4,0.949619) (0.5,0.948730)
};
\addlegendentry{Split}

\addplot+[crc] coordinates {
  (0.001,0.951511) (0.002,0.994596) (0.005,0.994580) (0.01,0.994603)
  (0.02,0.969884) (0.05,0.967724) (0.1,0.962343) (0.2,0.959218)
  (0.4,0.956721) (0.5,0.954822)
};
\addlegendentry{CRC}

\nextgroupplot[
  xlabel={Calibration ratio},
  ylabel={Avg. width [p.u.]},
  ymin=0.089, ymax=0.100,
  title={(b)}
]
\addplot+[split] coordinates {
  (0.001,0.090354) (0.002,0.093814) (0.005,0.092691) (0.01,0.092831)
  (0.02,0.093166) (0.05,0.093280) (0.1,0.093223) (0.2,0.093361)
  (0.4,0.093595) (0.5,0.093537)
};
\addlegendentry{Split}

\addplot+[crc] coordinates {
  (0.001,0.093814) (0.002,0.099311) (0.005,0.099311) (0.01,0.099311)
  (0.02,0.096194) (0.05,0.095858) (0.1,0.095192) (0.2,0.094800)
  (0.4,0.094449) (0.5,0.094305)
};
\addlegendentry{CRC}

\end{groupplot}

\node[anchor=north] at ($(group c1r1.south)!0.5!(group c2r1.south)-(0,1.0cm)$)
  {\pgfplotslegendfromname{grouplegend}};

\end{tikzpicture}